\documentclass[preprint]{vgtc}               

\graphicspath{{figures/}{pictures/}{images/}{./}} 

\usepackage{times}                     

\usepackage{tabu}                      
\usepackage{booktabs}                  
\usepackage{lipsum}                    
\usepackage{mwe}                       

\usepackage{mathptmx}                  

\vgtccategory{Poster}

\vgtcinsertpkg

\title{A Low-Latency 3D Live Remote Visualization System for Tourist Sites Integrating Dynamic and Pre-captured Static Point Clouds}

\author{
  Takahiro Matsumoto\thanks{e-mail: takahiro.matsumoto@ntt.com}%
  \and Masafumi Suzuki%
  \and Mariko Yamaguchi%
  \and Masakatsu Aoki%
  \and Shunsuke Konagai%
  \and Kazuhiko Murasaki%
}
\affiliation{\scriptsize NTT, Inc., Japan}

\teaser{
  \centering
  \includegraphics[width= \linewidth]{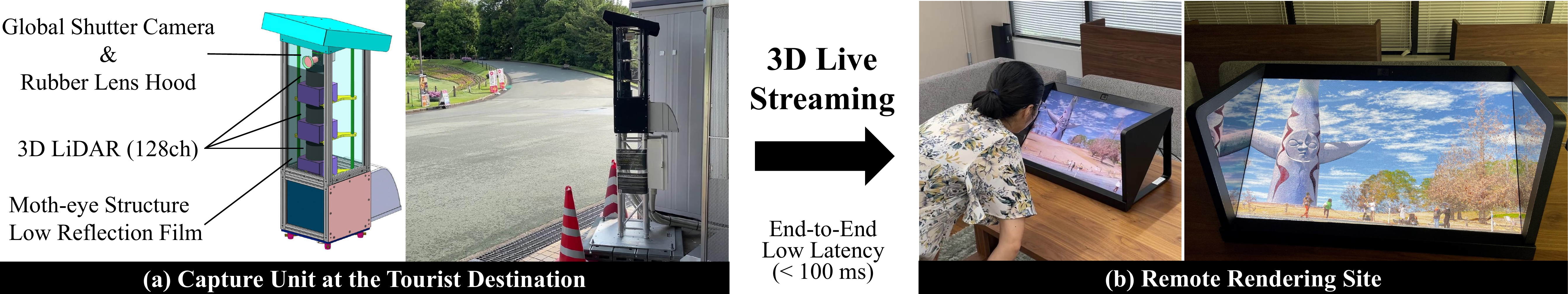}
  \caption{Overview of the proposed 3D live-visualization system. (a) Capture unit with protective housing installed at the tourist destination. (b) Remote rendering site displaying fused dynamic and static point clouds on a 3D display.}
  \label{fig1}
}

\abstract{
Various real-time methods for capturing and transmitting dynamic 3D spaces have been proposed, including those based on RGB-D cameras and volumetric capture. 
However, applying existing methods to outdoor tourist sites remains difficult because maintenance and aesthetic constraints limit sensor placement, and daylight variability complicates processing.
We propose a system that combines multiple LiDARs and cameras for live dynamic point cloud capture, and integrates them with pre-captured static point clouds for wide-area 3D visualization.
The system sustains 30 fps across wide-area scenes while keeping latency below 100 ms.
To mitigate lighting inconsistencies, static point-cloud colors are automatically adjusted to current lighting.
The effectiveness of our system is demonstrated through real-world deployment in a tourist site.
} 


\begin{document}


\firstsection{Introduction}
\maketitle
Live video streaming from tourist destinations is widely used to promote tourism, but current methods rely on conventional cameras and provide only 2D video, limiting spatial understanding and immersion for viewers.
In contrast, 3D streaming can offer greater presence, engagement \cite{Kuhne2023}, and improved spatial understanding \cite{Zheng2021}.

This study explores live 3D transmission over wide areas using few sensors while remaining robust to outdoor conditions (Figure~\ref{fig1}).
We propose a method that (i) captures and streams dynamic point clouds with multiple LiDARs and cameras, (ii) protects the sensors in a protective housing that resists weather and tampering, and (iii) fuses static and dynamic clouds by automatically re-tuning the static colors.
The effectiveness of the system is demonstrated through deployment at an actual tourist site, where end‑to‑end latency stays below 100 ms, LiDAR returns are preserved with moth‑eye films, and color consistency between static and dynamic point clouds is improved.

\vspace{-1mm}
\section{Related Work}
\vspace{-1mm}
Many approaches exist for static 3D capture, from SLAM and SfM to recent learning-based methods like NeRF and 3D Gaussian Splatting \cite{Christodoulides2025}.
However, these methods cannot capture changes in real time, such as the movement of people at tourist destinations.

Various methods have been proposed for dynamic 3D capture, among which RGB-D camera systems and volumetric capture rigs are prominent examples.
Yet, RGB-D cameras are limited to under 10 meters \cite{Xu2019}, and volumetric systems require many cameras or restrict the capture volume \cite{Suo2021}.
Consequently, these approaches remain unsuitable for wide outdoor environments with sparse installation points and variable lighting.

\vspace{-1mm}
\section{Proposed System}
\vspace{-1mm}
\subsection{Live Dynamic Point Cloud Transmission System}
We propose a real-time point-cloud transmission system for wide-area scenes that captures data using 128-channel rotating 3D LiDARs (1--200\,m, 10\,Hz) and global-shutter RGB cameras (Full HD, 30\,fps).
Both sensors are synchronized by Precision Time Protocol (PTP).
Each unit integrates three LiDARs and one camera. 
The LiDAR trio is synchronized over PTP with 120° phase offsets, and the camera is hardware-triggered whenever a LiDAR scan crosses its optical axis, yielding depth-color frames at 30 fps. 
A single unit already records 3D data in its vicinity, and deploying multiple units further enlarges the system’s spatial coverage.

However, three challenges remain: (i) \textbf{flicker}—the three LiDARs observe the same surface from slightly different angles, so static objects appear to jitter frame-to-frame; 
(ii) \textbf{occlusion miscoloring}—points seen by a LiDAR but hidden from the camera inherit incorrect RGB values; and 
(iii) \textbf{sparseness}—compared with the camera, the LiDAR’s far coarser sampling density limits our ability to obtain a high-density, colorized point cloud.

To overcome these issues, we developed three lightweight, real-time algorithms.
First, each LiDAR scan is projected onto its synchronized RGB frame, and motion masks derived from inter-frame RGB differences let us fuse only the static points of the latest three scans, eliminating flicker.
Second, depth consistency across the phase-shifted LiDARs is exploited to remove points that lie behind foreground surfaces, suppressing occlusion artifacts.
Finally, a joint bilateral filter guided by the RGB image upsamples the sparse depth to Full-HD resolution; we formulate this filter as a CNN and execute it in real time using PyTorch tensor operations \cite{Murasaki2025}.

\subsection{Protective Housing Design for Outdoor Deployment}
As shown in Figure~\ref{fig1}(a), the system is enclosed in acrylic panels that ensure safety while maintaining high transmittance in the visible band. 
Nevertheless, the panel surface reflects a portion of the LiDAR’s laser pulses, creating gaps in the captured point cloud. 
To mitigate this issue, we propose coating both faces of the acrylic with moth-eye anti-reflection films, which suppress these reflections and reduce point-cloud gaps.

\subsection{Color Transfer Method for Static Point Clouds}
To visualize a wider area, we choose to fuse a pre-captured static point cloud with live dynamic data.
However, the static cloud’s colors are fixed to the lighting conditions present at capture time; 
therefore, we adapt them to match the current illumination observed in the dynamic cloud.

We extend a color-transfer algorithm that aligns the color distribution of a source image with that of a target \cite{Reinhard2001}.
The original method converts the 2D image to CIELAB space, computes the per-channel mean and standard deviation, and shifts the target statistics to those of the source.

\begin{figure}[t] 
    \centering
    \includegraphics[width=85mm]{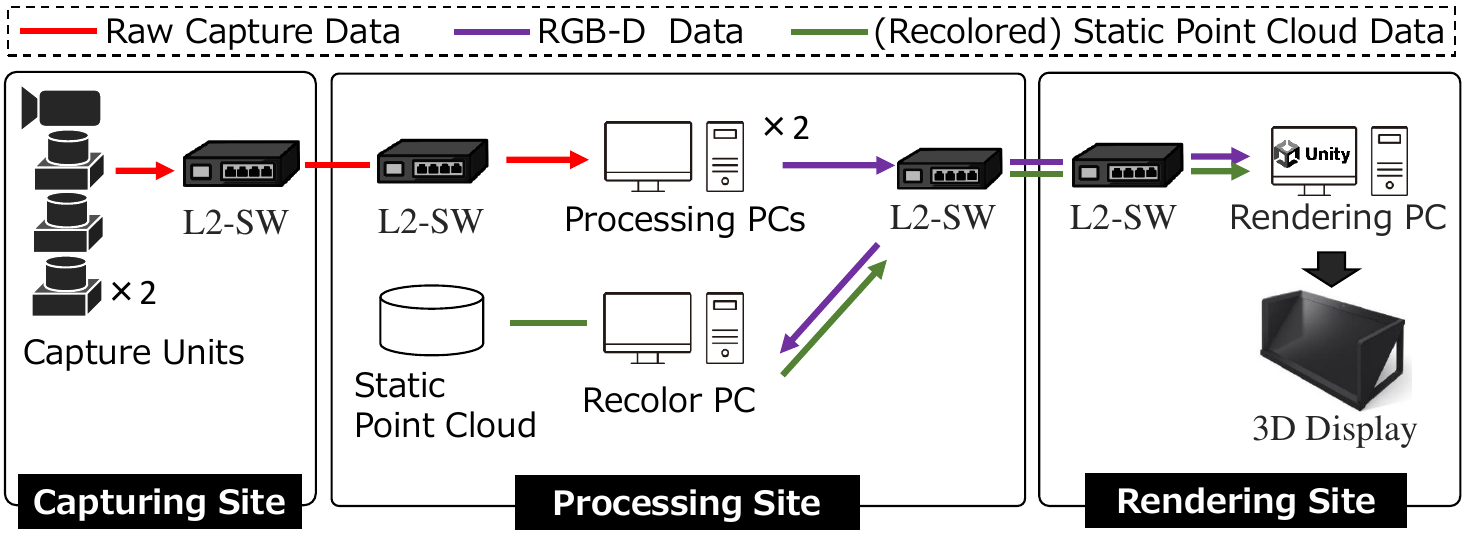}
    \caption{Overall configuration of the remote 3D live visualization system constructed at the tourist site.}
    \label{system_overview}
\end{figure}

We apply this algorithm to 3D point clouds with two key modifications.
First, color-transfer statistics are computed only from overlapping point pairs within a threshold distance l, because many points lie on moving objects or outside the static cloud’s footprint.
Second, to enhance local consistency, the static cloud is divided into k clusters using k-means, and color transfer is performed independently on each cluster.
The locally corrected colors are linearly blended with the global correction in a fixed ratio, achieving a balanced combination of global and local color adaptation.

\subsection{System Architecture}

Figure \ref{system_overview} shows the system setup.
Two capture units were installed, and data from each camera and LiDAR were sent to a processing site.
Two PCs processed the point-cloud data—performing densification, occlusion removal, and flicker suppression—and streamed compressed RGB-D frames to the visualization site at 30 fps.
RGB images were defocused to protect privacy.
At the visualization site, a rendering PC displays the RGB-D images on a 3D display.
Visualization runs in Unity and is displayed on a Sony ELF-SR2.
The processing site also has a Recolor PC, which updates static point-cloud colors every 15 minutes using the latest dynamic data.

For the field deployment, capture units were installed at a real tourist destination, while visualization was carried out at a remote site roughly 20 km away.
A pre-captured static point cloud was created by combining NavVis VLX 2 and Drone data of the park. 
The static cloud covers $6.64 \times 10^{4}\,m^{2}$ and has $3.88 \times 10^{7}$ points.

\vspace{-1mm}
\section{Experimental Results}
\vspace{-1mm}

\begin{figure}[tb] 
    \centering
    \includegraphics[width=85mm]{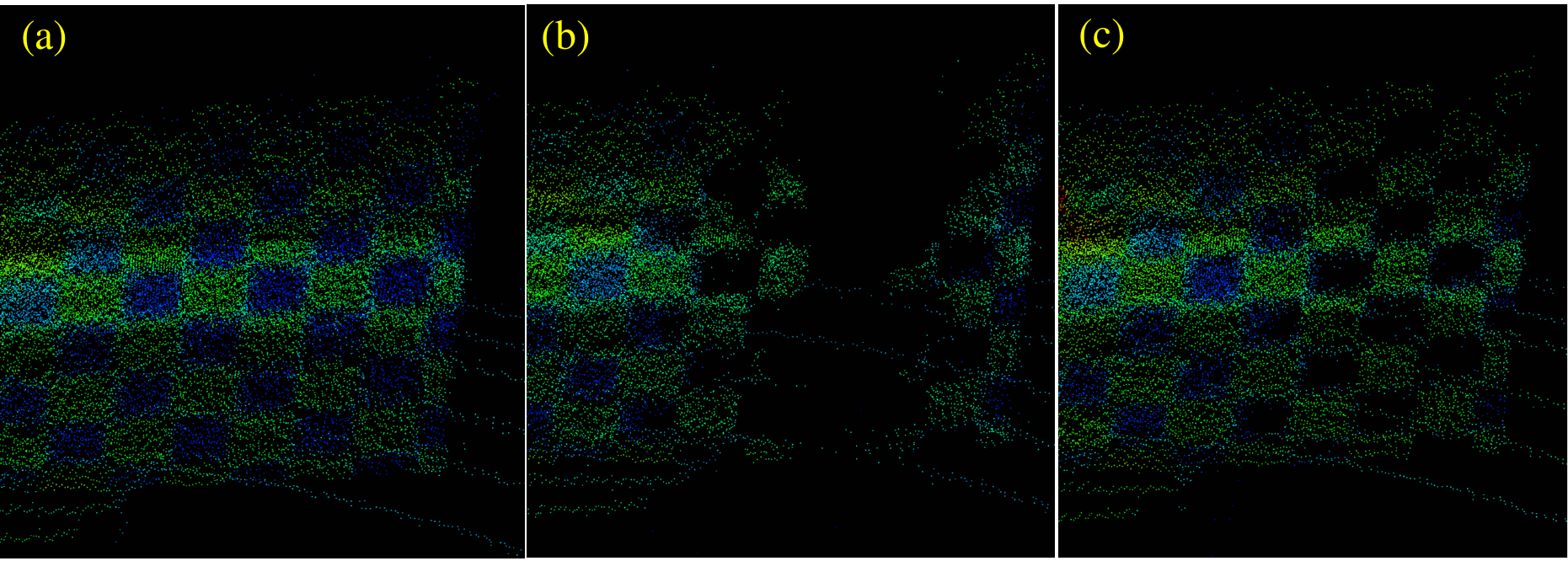}
    \caption{LiDAR returns under different housings—(a) no cover, (b) acrylic panel, (c) acrylic + moth-eye anti-reflection film.}   
    \label{moseye_film}
\end{figure}

We evaluated the dynamic point cloud system latency from LiDAR/camera capture to rendering, which averaged 81.31 ms (SD: 4.85), meeting real-time requirements.

Figure~\ref{moseye_film} quantifies how the moth-eye film suppresses LiDAR data loss. 
We evaluated three conditions—(a) no cover, (b) an acrylic panel, and (c) an acrylic panel coated on both faces with moth-eye film—using a 1.6 $\times$ 1.2\,m checkerboard (12\,cm squares) placed 1.5\,m from the LiDAR. 
Without any cover, the LiDAR captured \(7.14\times10^{4}\) points, whereas the acrylic panel alone caused a \(24.7\%\) loss.
In contrast, applying moth-eye films to both faces of the panel reduced the loss to just \(5.75\%\).

\begin{figure}[tb]
    \centering
    \includegraphics[width=85mm]{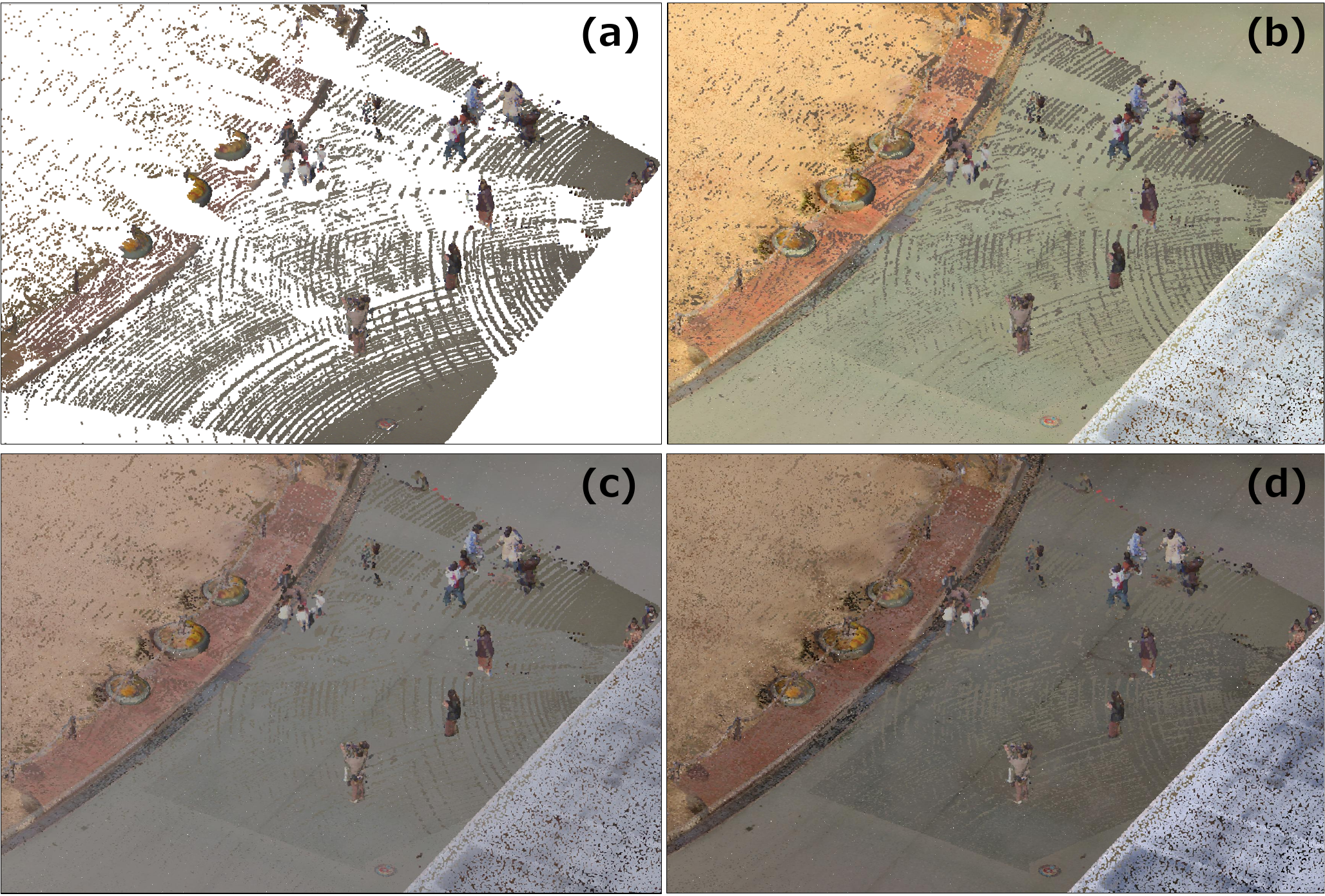}
    \caption{Color transfer results for static point cloud: (a) Reference dynamic point cloud, (b) Overlay of dynamic and static clouds, (c) Conventional method, (d) Proposed method.}
    \label{pc_color_transfer}
\end{figure}

Figure~\ref{pc_color_transfer} shows the color transfer results for the static point cloud. Figure~\ref{pc_color_transfer}(a) is a dynamic point cloud captured under cloudy conditions. In~\ref{pc_color_transfer}(b), simply overlaying the dynamic and static clouds reveals a clear color mismatch due to different sunlight conditions.
Figures~\ref{pc_color_transfer}(c) and (d) show the static point cloud after color transfer: (c) with the conventional method, (d) with ours. Our method achieves a closer color match to the dynamic point cloud.

Figure~\ref{fig1}(b) presents live visualization of the tourist site using both 2D video and a 3D display.
The 3D display overlays the live dynamic point cloud with the color-corrected static point cloud, enabling stereoscopic observation from arbitrary viewpoints, including overhead and tourist perspectives.
These results demonstrate the effectiveness of the system for live 3D remote visualization.

\vspace{-1mm}
\section{Conclusion}
\vspace{-1mm}
This study presents a system for transmitting and remotely visualizing 3D scenes of tourist sites by integrating live dynamic and static point clouds.
We have developed a capture and processing pipeline using multiple LiDARs and cameras, together with a protective housing that ensures reliable outdoor operation.
We have also introduced an adaptive color-adjustment technique for the static cloud, enabling robust visualization under varying lighting.

The capture units were deployed at a tourist site and demonstrated remote 3D visualization over 20 km. 
These results confirm the feasibility of our real-time 3D live-visualization system.

\bibliographystyle{abbrv-doi}
\bibliography{template}

\begin{thebibliography}{1}

\bibitem{Christodoulides2025}
A.~C. et~al.
\newblock Survey on 3d reconstruction techniques: Large-scale urban city reconstruction and requirements.
\newblock {\em IEEE Transactions on Visualization and Computer Graphics}, pp. 1--20, 2025.

\bibitem{Kuhne2023}
C.~K. et~al.
\newblock Direct comparison of virtual reality and 2d delivery on sense of presence, emotional and physiological outcome measures.
\newblock {\em Frontiers in Virtual Reality}, 4, 2023.

\bibitem{Reinhard2001}
E.~R. et~al.
\newblock Color transfer between images.
\newblock {\em IEEE Computer Graphics and Applications}, 21(5):34--41, 2001.

\bibitem{Murasaki2025}
K.~M. et~al.
\newblock Real-time lidar point cloud densification for low-latency spatial data transmission.
\newblock {\em IEICE Transactions on Information and Systems}, E109.D(3), 2026.

\bibitem{Xu2019}
L.~X. et~al.
\newblock Unstructuredfusion: Realtime 4d geometry and texture reconstruction using commercial rgb-d cameras.
\newblock {\em IEEE Transactions on Pattern Analysis and Machine Intelligence}, 42(10):2508--2522, 2019.

\bibitem{Zheng2021}
M.~Z. et~al.
\newblock How 2.5d maps design improve the wayfinding performance and spatial ability of map users.
\newblock {\em Informatics}, 2021.

\bibitem{Suo2021}
X.~S. et~al.
\newblock Neuralhumanfvv: Real-time neural volumetric human performance rendering using rgb cameras.
\newblock In {\em IEEE/CVF Conference on Computer Vision and Pattern Recognition}, 2021.

\end{thebibliography}
\end{document}